\documentclass[3p,times,procedia]{elsarticle}
\usepackage{nupha_ecrc}
\usepackage{placeins} % For FloatBarrier
\usepackage{float}
\usepackage{graphicx} % Include figure files
\usepackage{dcolumn} % Align table columns on decimal point
\usepackage{bm} % bold math
\usepackage{amsmath}
\usepackage[normalem]{ulem}
\usepackage{subcaption}
\volume{00}
\firstpage{1}

\journalname{Nuclear Physics A}

\runauth{}

\jid{nupha}

\jnltitlelogo{Nuclear Physics A}

\usepackage{amssymb}

\usepackage[figuresright]{rotating}

\newcommand{\ds}{\ensuremath{\displaystyle}}

\begin{document}

\begin{frontmatter}

%% Instructions from Editor: Please use the following \dochead only in the preprint version (e-print arXiv etc.); 
%% use empty \dochead{} when submitting to Nuclear Physics A!
\dochead{XXVIIth International Conference on Ultrarelativistic Nucleus-Nucleus Collisions\\ (Quark Matter 2018)}
%% e.g. \dochead{17th International Conference on Dynamical Processes in Excited States of Solids}

\title{The curvature of the chiral pseudocritical line from LQCD: analytic continuation and Taylor expansion compared.}


\author[infnpisa,difipi]{Claudio Bonati}
\author[infnpisa,difipi]{Massimo D'Elia}
\author[infnpisa]{Francesco Negro}
\author[infnroma]{Francesco Sanfilippo}
\author[parma]{Kevin Zambello}

\address[infnpisa]{INFN Sezione di Pisa, Largo B.~Pontecorvo 3, I-56127 Pisa, Italy}
\address[difipi]{Universit\`a di Pisa, Largo B.~Pontecorvo 3, I-56127 Pisa, Italy}
\address[infnroma]{INFN Sezione di Roma3, Via della Vasca Navale 84, I-00146 Roma, Italy}
\address[parma]{Universit\`a di Parma and INFN, Gruppo Collegato di Parma, Parco Area delle Scienze 7/A, I-43124 Parma, Italy}

\begin{abstract}
We present a determination of the curvature $\kappa$ of the chiral
pseudocritical line from $N_f=2+1$ lattice QCD at the physical point
obtained by adopting the Taylor expansion approach.
Numerical simulations performed at three lattice spacings
lead to a continuum extrapolated curvature
$\kappa = 0.0145(25)$, a value that is in excellent agreement with
continuum limit estimates obtained via analytic continuation
within the same discretization scheme,
$\kappa = 0.0135(20)$.
The agreement between the two calculations 
is a solid consistency check for both methods.
\end{abstract}


\end{frontmatter}

% \linenumbers


\section{Introduction}

Despite its great theoretical and phenomenological relevance, the temperature - baryon chemical potential ($T-\mu_B$)
QCD phase diagram is far from being fully understood.
Even first principle and non-perturbative approaches like Lattice QCD cannot directly access the $\mu_B>0$ region
because of the sign problem.
Anyhow, within the Lattice QCD framework, two methods have been quite extensively adopted to explore
the small $\mu_B$ part of the phase diagram: analytic continuation (AC) and Taylor expansion (TE).
We focus on the quadratic (in $\mu_B$) bending
of the pseudocritical line which separates the low-$T$ confined and chirally broken phase from the high-$T$ QGP phase.
The pseudocritical line can be parametrized as
\begin{equation}\label{corcur}
T_c(\mu_B)/T_c=1-\kappa\  (\mu_B/T_c)^2\, +\, O(\mu_B^4)\, ,
\end{equation}
where $\kappa$ is the curvature of the line.
In the literature, results obtained with the two methods,
even though the same discretization of QCD is adopted,
seem to indicate the presence of a tension: TE tends
to yield smaller values, about $\kappa\sim 0.006$ 
from Ref.~\cite{Endrodi2011},
as opposed to what is found via AC, i.e. about $\kappa\sim 0.014$
from Ref.~\cite{crow,corvo2}.
Such a tension is of more than two standard deviations.
The agreement between the methods would be necessary to be able to
state that all the possible systematics are under control.
We compare AC and TE adopting the same discretization to directly test such agreement (see Ref.~\cite{Bonati:2018nut} for more details).

\label{intro}
\section{Numerical setup and results}
\label{setupandres}
We adopted the same discretization setup that we used in our previous studies~\cite{crow,corvo2},
in which we discretized the $N_f=2+1$ QCD partition function $Z$
using the tree level Symanzik improved gauge action and the staggered stout smearing improved quark action. The gauge coupling and the bare quark masses are tuned to stay at the physical pion mass.

We based the determination of the crossover temperature on the study of the light quark condensate
$\langle\bar\psi\psi\rangle_l=(T/V)\cdot \partial \log Z/\partial m_l=
\langle\bar{u}u\rangle+\langle\bar{d}d\rangle\ $,
where $V$ is the spatial volume and $m_l$ is the bare light quark mass.
Two different prescriptions have been adopted to handle the additive and multiplicative renormalization this observable is affected by
\begin{equation} \label{rencond1e2}
\langle\bar{\psi}\psi\rangle_{r1}\,(T)\equiv\frac{\left[
\langle \bar{\psi}\psi\rangle_l -\frac{\ds 2m_{l}}{\ds m_s}\langle \bar{s}s\rangle\right](T)}{\left[\langle \bar{\psi}\psi\rangle_l-\frac{\ds 2m_{l}}{\ds m_s}\langle \bar{s}s\rangle\right](T=0)}\ \ \ {\textnormal{and}} \ \ \ \langle \bar{\psi}\psi\rangle_{r2}\,(T)=\frac{m_{l}}{m_{\pi}^4}\left(\langle\bar{\psi}\psi\rangle_{l}\,(T) -\langle\bar{\psi}\psi\rangle_{l}\,(T=0)\right)\,,
\end{equation}
introduced respectively in Ref.~\cite{Cheng:2007jq} and Ref.~\cite{Endrodi2011} (the $\mu_B$ dependence of finite $T$ observables is understood). This allowed us to check for possible systematics related to the specific renormalization prescription.
\subsection{Analytic continuation approach}
We briefly summarize the results we obtained in Ref.~\cite{crow,corvo2} within the analytic continuation framework.
We performed numerical simulations at non zero imaginary baryon chemical potential ($\mu_{B,I} = i \mu_B$), which is a sign problem free theory. 
We adopted a setup with degenerate light quark chemical potentials  $\mu_{l,I}/(\pi T)=\mu_{B,I}/(3\pi T)$
and zero strange quark chemical potential $\mu_{s,I}=0$.
At fixed $\mu_{l,I}/(\pi T)$ chosen in the set $\lbrace0,0.2,0.24,0.275\rbrace$, we computed $\langle\overline{\psi}\psi \rangle_r$ and the renormalized chiral susceptibility $\chi_{\overline\psi\psi}^r$
at several temperatures around $T_c$.
\begin{figure}[b!]
\caption{}
\vspace{-0.3cm}
\begin{subfigure}{.333\textwidth}
  \centering
  \includegraphics[width=.95\linewidth]{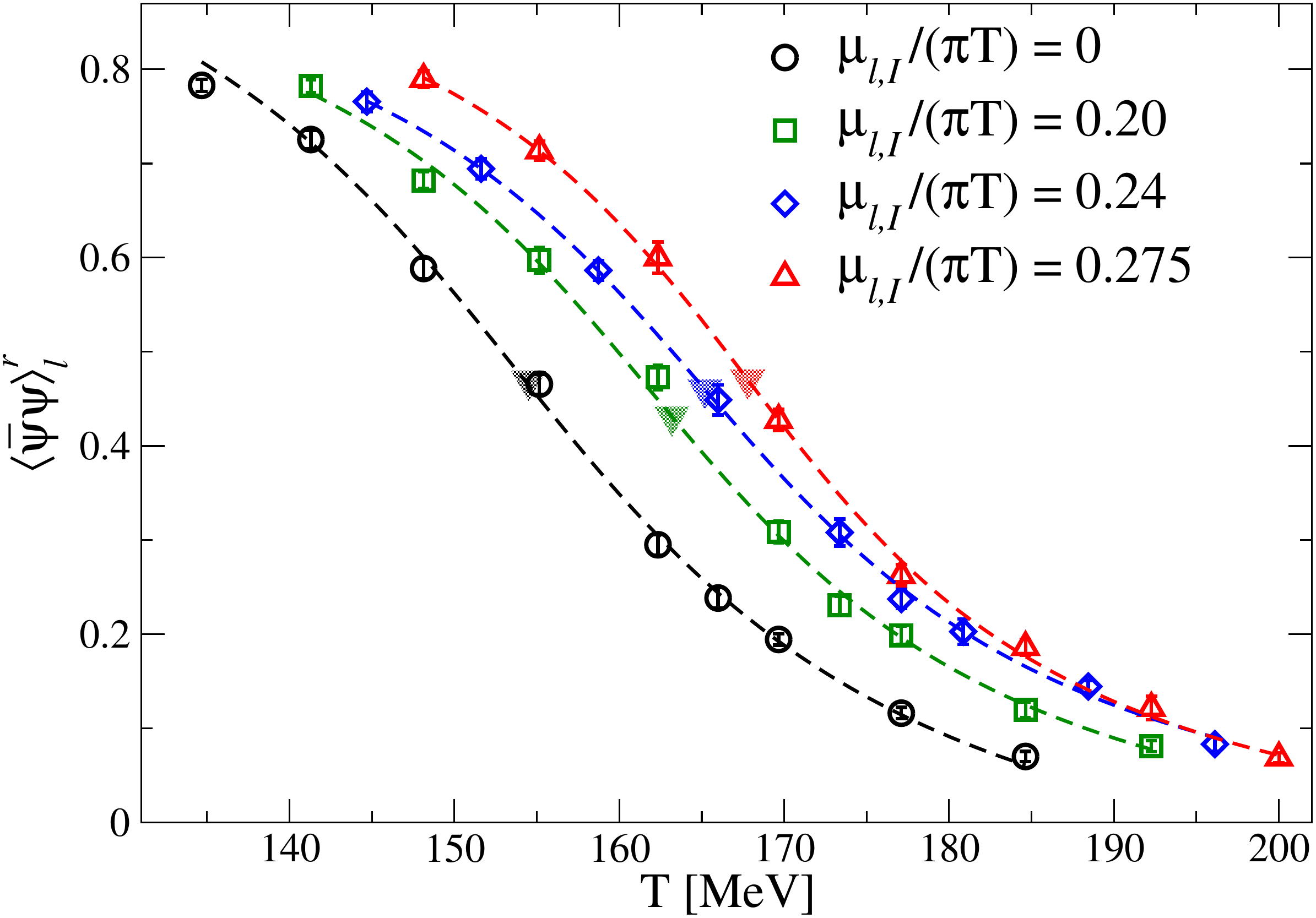}
  \label{fig:sfig1}
\end{subfigure}%
\begin{subfigure}{.333\textwidth}
  \centering
\vspace{0.1cm}  \includegraphics[width=.94\linewidth]{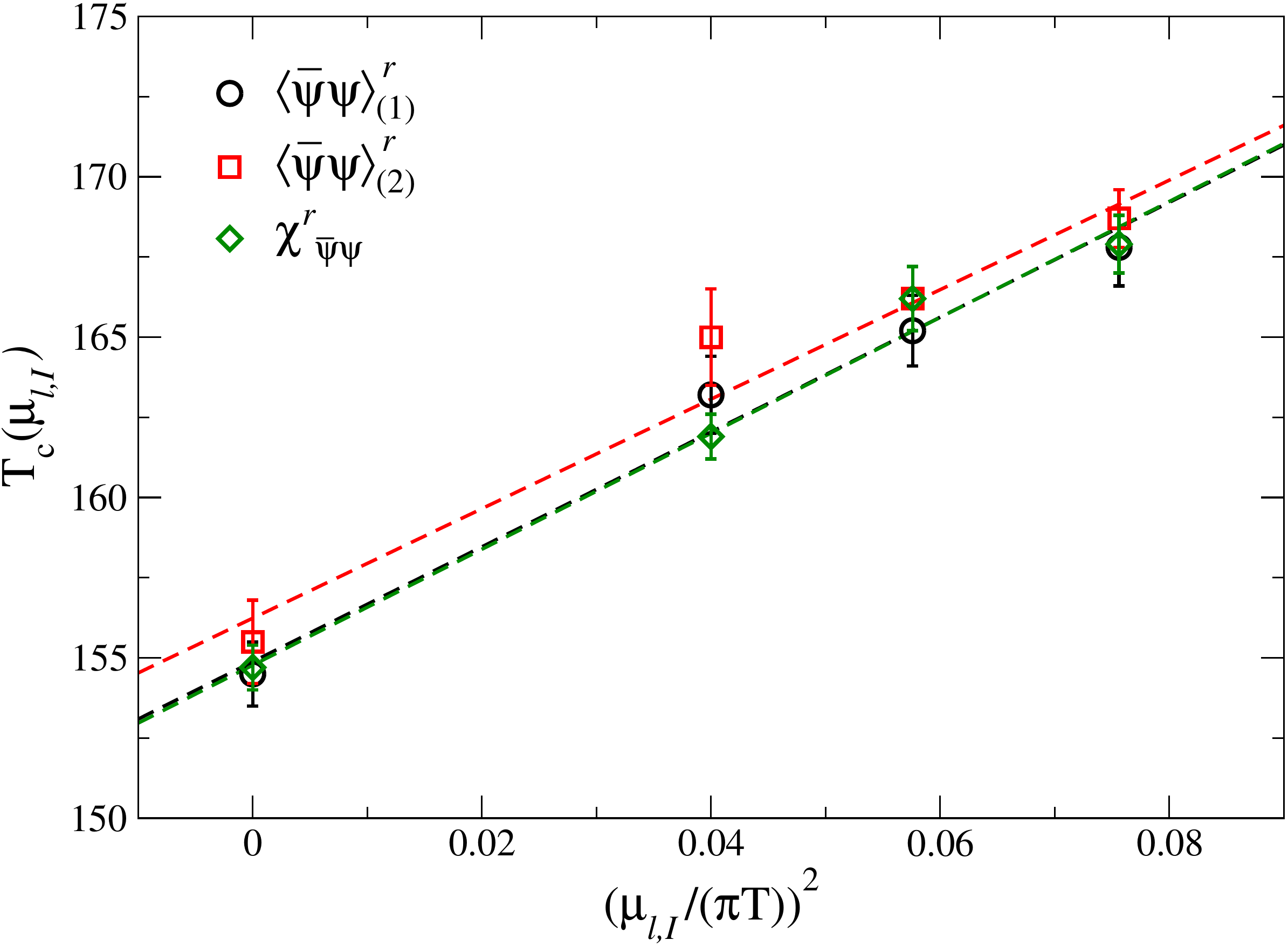}
  \label{fig:sfig2}
\end{subfigure}
\begin{subfigure}{.333\textwidth}
  \centering
  \includegraphics[width=.95\linewidth]{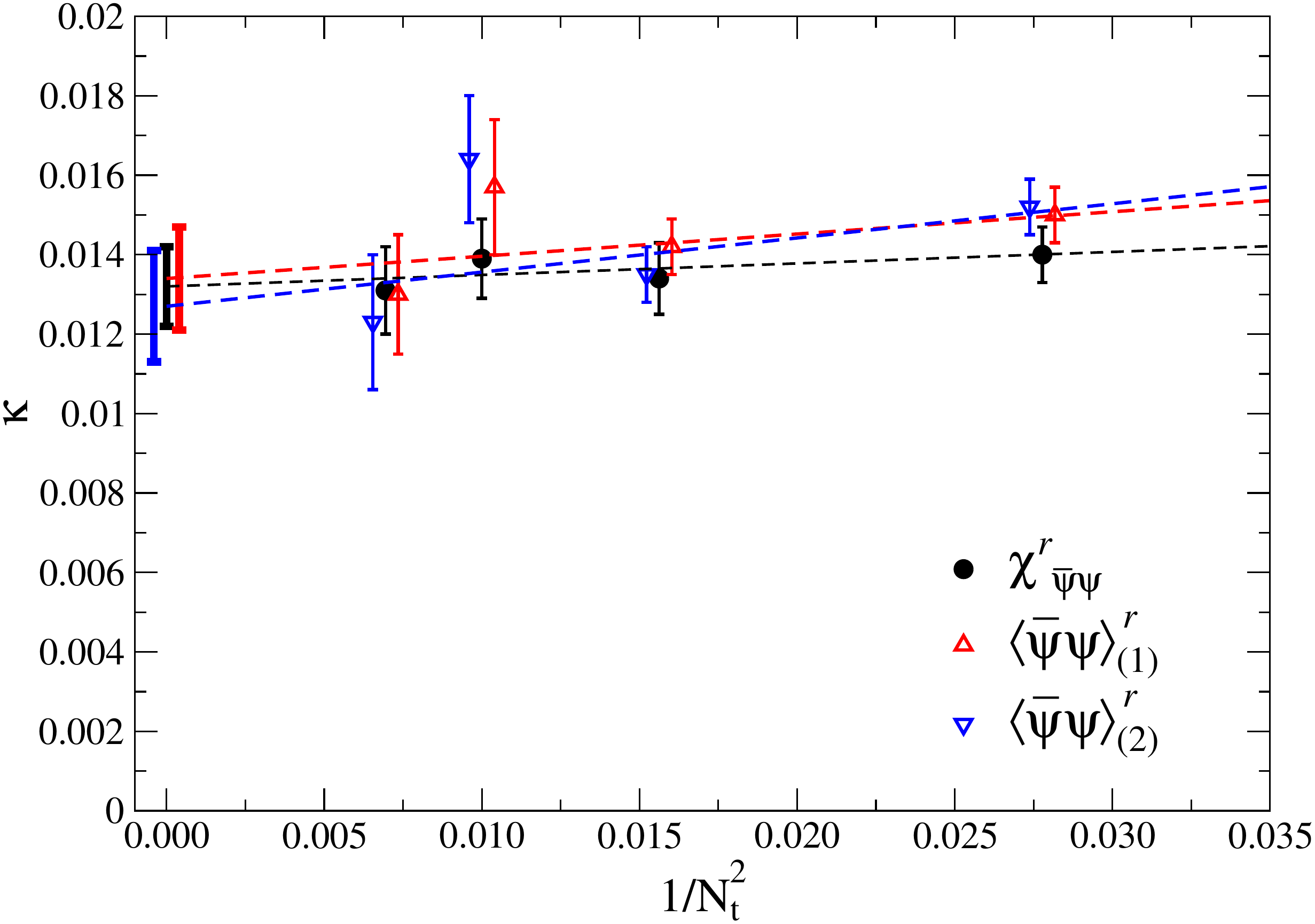}
  \label{fig:sfig3}
\end{subfigure}
\label{fig:fig1}
\end{figure}
The crossover temperature has been estimated by identifying the inflection point of the condensate (see {\it left panel} of Fig.\ref{fig:fig1}) and the peak of its susceptibility for all the explored values of $\mu_{l,I}/(\pi T)$.
As shown in the {\it center panel} of Fig.~\ref{fig:fig1}, we extracted the curvature $\kappa$
by fitting the data of $T_c(\mu_{l,I}/(\pi T))$ with a linear function in $(\mu_{l,I}/(\pi T))^2$.
The same procedure has been repeated on lattices with $N_t=6,\,8,\,10,\,12$ and then the
continuum limit for $\kappa$ taken assuming $O(1/N_t^2)$ finite lattice spacing effects ({\it right panel} of Fig.\ref{fig:fig1}).
Our final estimate, which takes into account several possible systematics (among which a procedure to approach the continuum limit in a different way from that described here) is $\kappa=0.0135(20)$.

\subsection{Taylor expansion approach}
\begin{figure}[h!]
\caption{}
\vspace{-0.7cm}
\begin{subfigure}{.333\textwidth}
  \centering
  \includegraphics[width=1.03\linewidth]{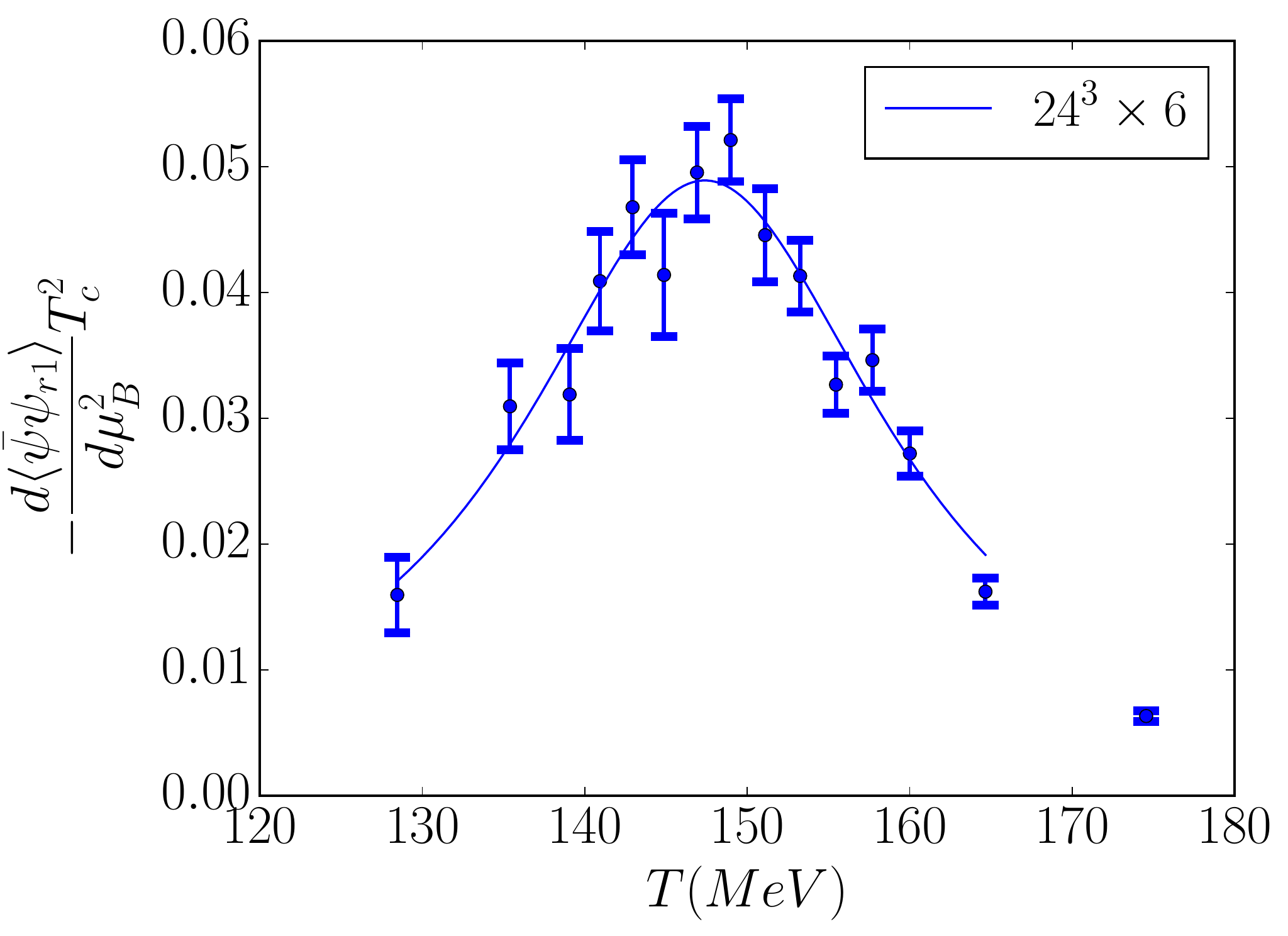}
\end{subfigure}
\begin{subfigure}{.333\textwidth}
  \centering
\vspace{0.3cm}
  \includegraphics[width=1.00\linewidth]{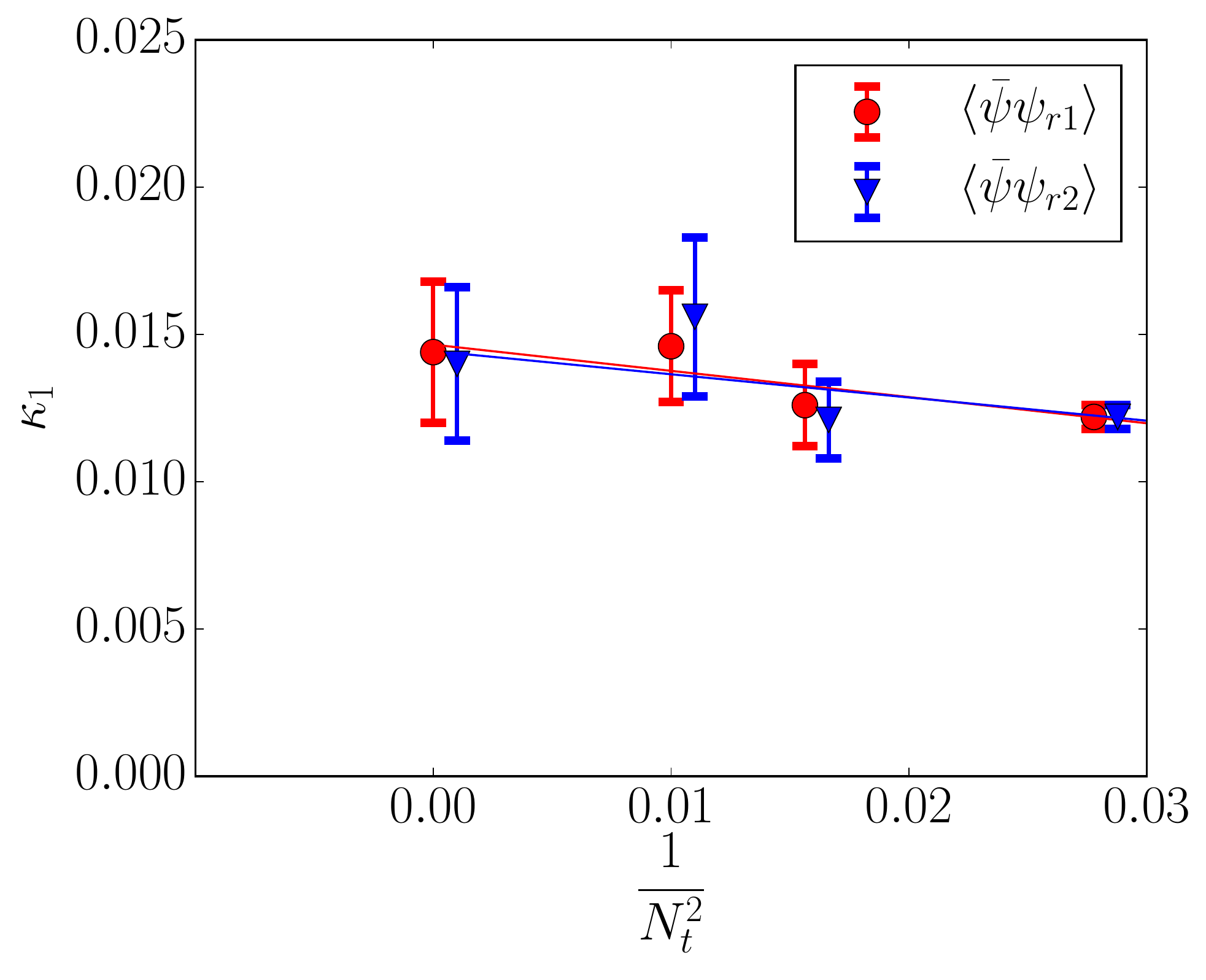}
\end{subfigure}
\begin{subfigure}{.333\textwidth}
  \centering
\vspace{-0.1cm}
  \includegraphics[width=1.10\linewidth]{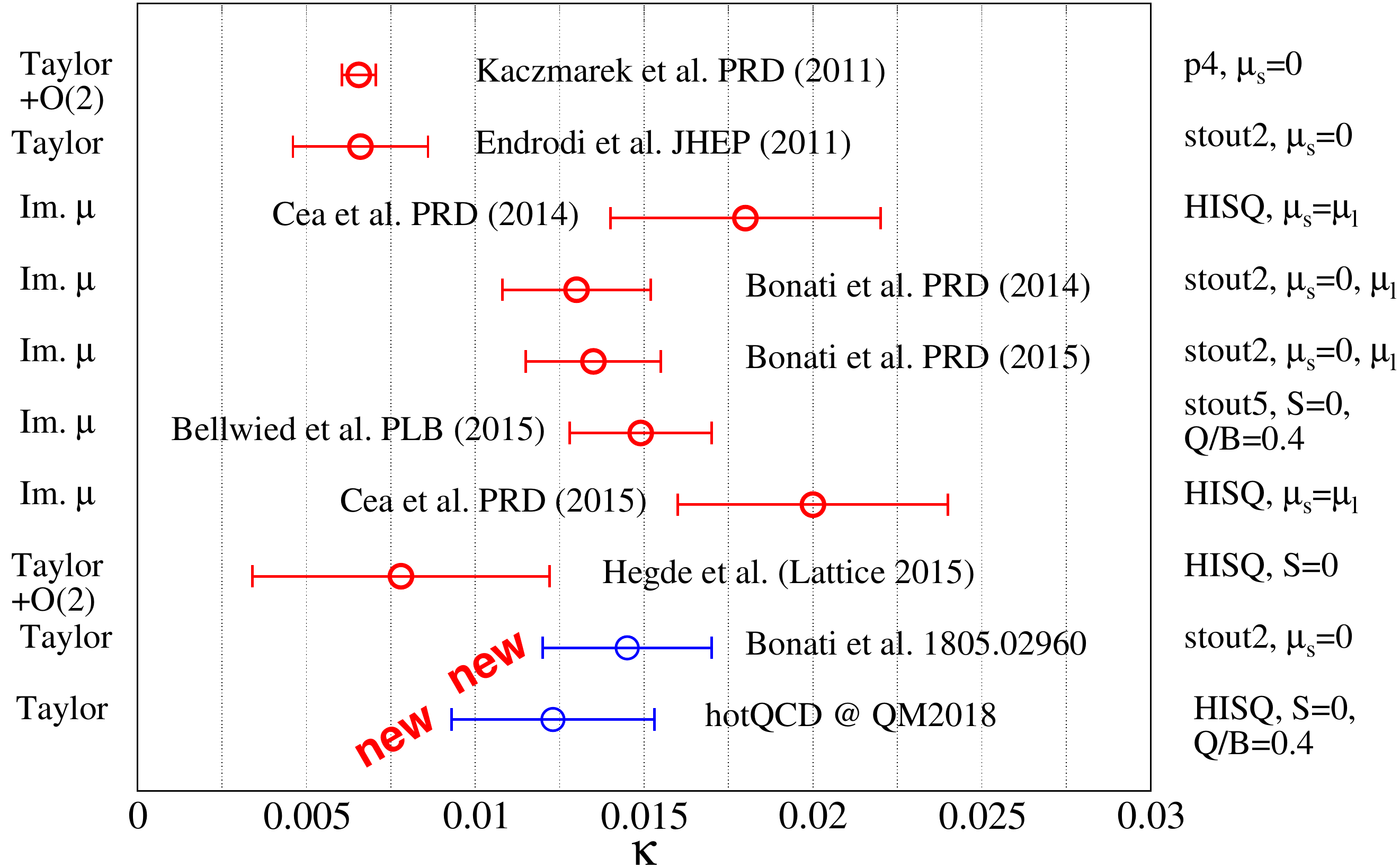}
\end{subfigure}
\label{fig:fig2}
\vspace{-1.3cm}
\end{figure}

At $\mu_B=0$, we define the crossover temperature $T_c$ as the inflection point of the renormalized quark condensate.
To investigate the small $\mu_B$ region, we consider the Taylor expansion of $\langle\bar{\psi}\psi\rangle_r$ to order $\mu_B^2$
\begin{equation}
\langle \bar{\psi} \psi \rangle_r\,(T, \mu_B) = \langle \bar{\psi} \psi \rangle_r\,(T, 0) + \mu_B^2 \frac{\partial \langle \bar{\psi} \psi \rangle_r}{\partial ( \mu_B^2 )} (T,0) + O(\mu_B^4).
\end{equation}
It is quite natural to extend the definition of $T_c$ 
by looking for an inflection point at fixed $\mu_B\neq 0$:
a point where
$\partial^2 \langle\bar{\psi}\psi\rangle^r(T,\mu_B)/ \partial T^2 = 0$.
Since the formula for $\kappa$ that can be derived from this equation proved to be highly numerically demanding,
we considered also an alternative prescription for $\kappa$ (introduced in Ref.~\cite{Endrodi2011}),
where the pseudo-critical temperature at $\mu_B\neq 0$ is defined as the temperature
where the renormalized condensate remains at the same value as at $T_c$ for $\mu_B  = 0$: $ \langle \bar{\psi}\psi \rangle^r (T, \mu_B^2)|_{T=T_c(\mu_B^2)}  \equiv \langle \bar{\psi}\psi \rangle^r (T_c, 0)$.
The formulas for $\kappa$, that can be obtained by imposing these two conditions, read respectively
\begin{equation}
\label{defkappa12}
    \kappa_{inflection} = T_c\frac{   \frac{\partial^2}{\partial T^2} ( \frac{\partial \langle \bar{\psi} \psi \rangle^r(T,\mu_B)}{\partial (\mu_B^2)}|_{\mu_B=0} ) |_{T=T_c}   }{   \frac{\partial^3}{\partial T^3} \langle \bar{\psi} \psi \rangle^r(T,0) |_{T=T_c}   } \ \ \ {\textnormal{and}}\ \ \ \kappa_{fixed\ \langle\overline\psi\psi\rangle_r} = T_c \frac{ \frac{\partial \langle \bar{\psi}\psi \rangle^r}{\partial (\mu_B^2)}|_{\mu_B = 0, T=T_c} }{ \frac{\partial \langle \bar{\psi}\psi \rangle^r}{\partial T}|_{\mu_B = 0, T = T_c} } \mbox{ . }
\end{equation}
We performed simulations on $24^3\times 6$, $32^3\times 8$ and $40^3\times 10$ lattices
at several temperatures and we estimated the $T-$derivatives appearing in Eq.~(\ref{defkappa12}) by
fitting data with suitable functions: $atan$, $tanh$ and a cubic polynomial for the condensate and a lorentzian function, a quadratic polynomial and a cubic spline for its $\mu_B^2$ derivative.
An example is reported in the {\it left panel} of Fig.~\ref{fig:fig2} for the $\mu_B^2$ derivative of the condensate.
Even on our coarsest lattice, statistical uncertainty for $\kappa_{inflection}$ is very large, about $30\%$; anyhow the estimate we get
is compatible with $\kappa_{fixed\ \langle\overline\psi\psi\rangle_r}$.
With the present statistics data for $\kappa_{fixed\ \langle\overline\psi\psi\rangle_r}$ are, on the other hand, 
precise enough to perform the continuum limit extrapolation (see {\it center panel} of Fig.~\ref{fig:fig2}), leading to the continuum
estimate $\kappa=0.0145(25)$.

\vspace{-0.1cm}
\section{Conclusions}
\label{conc}
As anticipated,
the results we find with the two approaches agree:
$\kappa=0.0135(20)$ for analytic continuation and
$\kappa=0.0145(25)$ for Taylor expansion.
Also recent results by the HotQCD collaboration
(as presented by P. Steinbrecher at the Quark Matter 2018 conference)
indicate that the tension is getting solved. We report in the {\it right panel}
of Fig.~\ref{fig:fig2} an updated summary plot of the determinations of $\kappa$
in the literature from Ref.~\cite{Kaczmarek2011,Endrodi2011,ccp,crow,corvo2,ntc,ccp2,hegde,Bonati:2018nut}
and from the mentioned talk by P. Steinbrecher for the HotQCD collaboration at QM2018.

\vspace{-0.3cm}
\section*{Acknowledgement}
FN acknowledges financial support from the INFN HPC\_HTC project.



%% References


%\bibliographystyle{elsarticle-num}



\end{document}